\documentclass[aps,superscriptaddress,amsmath,amssymb,twocolumn,showpacs,floatfix,english,noshowpacs]{revtex4-2}

\usepackage{url}
\usepackage{amsmath}
\usepackage{amssymb}
\usepackage{bm}
\usepackage{bbold}
\usepackage{graphicx}
\usepackage{xcolor}

\usepackage[colorlinks=true, urlcolor=blue, linkcolor=blue, citecolor=blue, pdftex]{hyperref}
\usepackage[english]{babel}

\usepackage{soul}
\usepackage{blindtext}
\usepackage{lipsum}
\usepackage{nicefrac}
\usepackage{cancel}

\newcommand{\dagga}{{\phantom{\dagger}}}

\begin{document}

\title{A many-body marker for three-dimensional topological insulators with inversion symmetry}

\author{Federico Becca}
\affiliation{Dipartimento di Fisica, Universit\`a di Trieste, Strada Costiera 11, I-34151 Trieste, Italy}
\author{Alberto Parola}
\affiliation{Dipartimento di Scienza e Alta Tecnologia, Universit\`a dell'Insubria, Via Valleggio 11, I-22100 Como, Italy}

\date{\today}

\begin{abstract}
We extend the previously defined many-body marker for two-dimensional $\mathbb{Z}_2$ topological insulators [I. Gilardoni {\it et al.}, Phys. Rev. B {\bf 106}, 
L161106 (2022)] to distinguish trivial, weak-, and strong-topological insulators in three dimensions, in the presence of inversion symmetry. The marker is 
written in term of ground-state expectation values of position operators and can be employed to detect topological phases of interacting systems beyond 
mean-field approximations, e.g., within quantum Monte Carlo techniques. Here, we show that the correct results of the non-interacting limit are reproduced 
by the many-body marker.
\end{abstract}

\maketitle

\section{Introduction}\label{sec:intro}

The groundbreaking proposals~\cite{kane2005a,bernevig2006} that established the existence of topological insulators in two spatial dimensions (resulting from 
spin-orbit coupling and time-reversal invariance) and their subsequent experimental confirmation~\cite{konig2007} ignited a tremendous amount of activity in 
this area~\cite{hasan2010}. A remarkable achievement was the generalization to three-dimensional systems~\cite{fu2007a,moore2007,roy2009}. Classifications of 
insulators have been developed within the non-interacting framework~\cite{chiu2016}, emphasizing the role of specific symmetries in defining distinct topological 
states. In particular, when only time-reversal and charge conjugation symmetries are allowed, ten classes were identified and, in each spatial dimension, 
five of them are topological, while the remaining ones are trivial. In general, the fingerprint of topological states is the existence of gapless excitations 
at the boundaries, typically referred to as edge states. In addition, the bulk of the system can be characterized by topological invariants (or markers) that, 
in the simplest case are given by Chern or Chern-Simons forms (in even or odd dimensions, respectively). In two dimensions, the most celebrated example is 
given by integer quantum Hall states, which can be stabilized even in the absence of specific symmetries. In this context, the Chern number is related to the 
(quantized) transverse conductivity~\cite{thouless1982}. Consequently, an infinite number of topologically different states can exist, stemming from the fact 
that the Chern number can assume any integer value $\mathbb{Z}$. Much of our understanding in this field has been obtained from studying systems within the 
band theory paradigm. Here, the Chern number can be computed by integrating the Berry curvature over the first Brillouin zone~\cite{thouless1982,haldane1988}. 
The situation is radically different in three dimensions, where quantum Hall states do not exist. The presence of time-reversal symmetry alone fundamentally 
changes the situation limiting the possible states to just two classes (trivial and topological) in both two and three dimensions, leading to a $\mathbb{Z}_2$ 
invariant~\cite{chiu2016}. Nevertheless, the actual computation of topological markers is more involved than that of Chern numbers and different approaches 
have been suggested~\cite{fu2006}. The presence of additional (lattice) symmetries greatly simplifies the situation, as emphasized for the case in which the 
inversion symmetry is present~\cite{fu2007b}. Then, the classification does not change in two dimensions, still giving rise to a single $\mathbb{Z}_2$ invariant. 
Instead, in three dimensions, four $\mathbb{Z}_2$ invariants emerge, distinguishing the trivial insulator from the so-called weak- and strong-topological 
insulators\cite{fu2007a,moore2007,roy2009}. All these invariants can be constructed from the parity eigenvalues of the occupied orbitals at time-reversal 
momenta in the Brillouin zone~\cite{fu2007b}.

The most significant limitation of previous attempts to define topological markers is their reliance on non-interacting systems, where a single-particle 
picture is used. The inherent presence of interactions in all states of matter, including band insulators, necessitates a general definition of topological 
insulators that accounts for electron-electron interactions. The importance of this requirement is highlighted by the one-dimensional example, which 
demonstrates that a non-interacting topological state may be unstable against interactions~\cite{fidkowski2010}. Interestingly, a generalization of the
results obtained in Ref.~\cite{thouless1982} to include many-body effects has been proposed for (two-dimensional) quantum Hall states~\cite{niu1985}.
However, this approach necessitates the computation of the ground-state wave function with different (twisted) boundary conditions, posing challenges 
for numerical implementations~\cite{kudo2019}. Building upon Volovik's initial observation~\cite{volovik1988,volovik2009}, it has been proposed that 
topological aspects can be identified within the (interacting) single-particle Green's function $G({\bf k},\omega)$~\cite{gurarie2011,blason2023}, where 
edge poles might become zeros without the single-particle gap closing. This possibility has been explored in order to give a topological characterization
of the Mott phase of lattice models~\cite{wagner2023}. Furthermore, it has been argued that an effective Hamiltonian can be defined from $G({\bf k},0)$, 
allowing the formulation of topological invariants for interacting systems~\cite{wang2012}. A few works applied these ideas to the Bernevig-Hughes-Zhang 
(BHZ)~\cite{amaricci2016,amaricci2017} and Hofstadter models~\cite{irsigler2019} in the presence of electron-electron repulsion. Still, this kind of 
approaches is not always achievable, since the Green's function is not easily assessed in several numerical methods (e.g., quantum Monte Carlo).

From a different perspective, starting from the position operator introduced within the modern theory of polarization of insulators~\cite{resta1999}
(and its generalization to include conducting phases~\cite{aligia1999}), we have shown that it is possible to define a $\mathbb{Z}_2$ many-body marker, 
whose sign discriminates between trivial and topological insulators, for both one and two spatial dimensions~\cite{gilardoni2022}. Here, the marker 
is expressed as a ground-state expectation value of a suitably defined many-body operator, allowing for straightforward evaluation within various 
variational approaches. A first accomplishment of this strategy has been obtained within the one-dimensional BHZ model with intra-orbital Hubbard 
interaction~\cite{barbarino2019}, where the stability of trivial and topological phases has been accurately determined by using Jastrow-Slater wave 
functions and density-matrix renormalization group techniques~\cite{favata2025}.

Here, we further elaborate along the same direction and show that our many-body marker can be also used in three dimensions with inversion symmetry, 
to distinguish among trivial, weak-, and strong-topological insulators. The picture becomes particularly transparent within the non-interacting limit 
for models with conserved $z$-component of the spin (i.e., whenever the Hamiltonian commutes with $S^z$). In this case, the sign of our many-body marker 
is fully determined by the signs of the Chern numbers within two planes in the Brillouin zone. This outcome is reminiscent of the fact that the integral 
of the (three-dimensional) Chern-Simons form equals the parity of the sum of the Chern numbers computed at two fixed planes in the $k$-space (e.g., 
$k_z=0$ and $\pi$ on the cubic lattice)~\cite{wang2010}. In addition, the marker is also simply related to the parity eigenvalues of the occupied 
orbitals at time-reversal invariant momenta (TRIM). A similar structure is also obtained, within the non-interacting limit, for general models that 
do not commute with $S^z$. Our topological marker allows one to consider interacting models by using variational or projection quantum Monte Carlo 
techniques. These studies require heavy numerical calculations and, therefore, are left for future investigations.

The paper is organized as follows: in section~\ref{sec:general} we give the general definitions for the lattice structure and the properties of the
Hamiltonian; in section~\ref{sec:marker}, we define the topological marker, discussing its expression in the non-interacting limit; in section~\ref{sec:models},
we report the calculations for two specific models that have been introduced in the past; finally, in section~\ref{sec:concl}, we draw our conclusions.

\section{General aspects}\label{sec:general}

A $d$-dimensional Bravais lattice is defined by a set of $d$ primitive vectors ${\bf a}_i$, such that ${\bf R}=\sum_{j} l_j {\bf a}_j$, where $l_j$ are 
integers. The vectors in the reciprocal space are defined by the usual condition ${\bf a}_i \cdot {\bf b}_j = 2\pi \delta_{ij}$; the reciprocal-lattice
vectors are then given by ${\bf G}=\sum_{j} m_j {\bf b}_j$, where $m_j$ are integers. The geometry of a finite cluster with periodic boundary conditions 
is identified by $d$ translation vectors ${\bf T}_i= \sum_{j} N_{ij} {\bf a}_j$ (where $N_{ij}$ is a matrix of integers) and the quantization rule in the 
Brillouin zone is simply $x_i = {\bf q} \cdot {\bf a}_i = 2\pi \sum_{j} (N^{-1})_{ij} n_j$ (where $n_i$ are integers). In this work, we focus our attention 
on $d=3$ and take clusters defined by a diagonal matrix with $N_{11}=L_1$, $N_{22}=L_2$, and $N_{33}=L_3$. The total number of unit cells is 
$N=L_1 \times L_2 \times L_3$. In this case, we have that $x_i=2\pi n_i/L_i$, with $n_i=0,\dots,L_i-1$ for $i=1,2,3$. Setting ${\bf q}=\sum_i q_i\,{\bf b}_i$, 
the quantization simply gives $q_i=n_i/L_i$.

Let us consider a generic non-interacting and translationally invariant model, with $N_{\rm o}$ orbitals per primitive cell, in a cluster with $N_{\rm c}$ cells 
and periodic-boundary conditions. Then, the Hamiltonian can be written in Fourier space in terms of a $k$-dependent $2 N_{\rm o}\times 2 N_{\rm o}$ matrix 
$h({\bf k})$ (that includes both orbital and spin degrees of freedom):
\begin{equation}\label{eq:hamiltonian}
\hat{\cal H} = \sum_{\bf k} \sum_{\eta,\sigma,\eta^\prime,\sigma^\prime} h^{\sigma,\sigma^\prime}_{\eta,\eta^\prime}({\bf k}) \, \hat{c}^\dagger_{{\bf k},\eta,\sigma} 
\hat{c}^\dagga_{{\bf k},\eta^\prime,\sigma^\prime},
\end{equation}
where $\hat{c}^\dagger_{{\bf k},\eta,\sigma}$ ($\hat{c}^\dagga_{{\bf k},\eta,\sigma}$) is the fermionic creator (annihilator) operator of an electron of momentum 
${\bf k}$, orbital $\eta$ and spin $\sigma$. On any Bravais lattice, these fermionic operators are the Fourier transformed of $\hat{c}^\dagger_{{\bf R},\eta,\sigma}$ 
and $\hat{c}^\dagga_{{\bf R},\eta,\sigma}$, which create and destroy an electron on the site ${\bf R}$, orbital $\eta$, and spin $\sigma$.

The Hamiltonian~\eqref{eq:hamiltonian} can be easily diagonalized by a canonical transformation:
\begin{equation}
\hat{\phi}^\dagger_{{\bf k},a} = \sum_{\eta,\sigma} u_{{\bf k},a}(\eta,\sigma) \hat{c}^\dagger_{{\bf k},\eta,\sigma},
\end{equation}
where $a=1,\dots,2 N_{\rm o}$ indicates the band index and $u_{{\bf k},a}$ are the eigenvectors of the matrix $h({\bf k})$. Then, the Bloch functions are defined by
\begin{equation}\label{eq:bloch}
\psi_{{\bf k},a}({\bf R},\eta,\sigma) = \langle 0| \hat{c}^\dagga_{{\bf R},\eta,\sigma} \hat{\phi}^\dagger_{{\bf k},a} |0\rangle = \frac{e^{i{\bf k} \cdot {\bf R}}}
{\sqrt{N_{\rm c}}} u_{{\bf k},a}(\eta,\sigma).
\end{equation}
It is convenient to express the $2 N_{\rm o} \times 2 N_{\rm o}$ matrix $h({\bf k})$ in terms of a tensor product of $2 \times 2$ Pauli matrices $\sigma_{\alpha}$ 
(with $\alpha=0,\dots,3$, including $\sigma_{0}=\mathbb{1}$) and $N_{\rm o} \times N_{\rm o}$ matrices (that, in the simplest case with $N_{\rm o}=2$ are also Pauli 
matrices $\tau_{\beta}$), which act on the spin and orbital indices, respectively. Here, we focus on the cases in which the Hamiltonian $\hat{\cal H}$ is invariant 
under inversion and time-reversal symmetries. The (unitary) inversion operator $\hat{\cal P}$ transforms the fermionic operator as 
\begin{equation}
\hat{\cal P}^{-1} \,\hat{c}^\dagga_{{\bf R},\eta,\sigma} \,\hat{\cal P} = \sum_{\eta^\prime} V_{\eta,\eta^\prime} \hat{c}^\dagga_{-{\bf R},\eta^\prime,\sigma},
\end{equation}
where $V$ is a $N_{\rm o} \times N_{\rm o}$ (model-dependent) unitary matrix acting on band indices. Instead, the (anti-unitary) time-reversal operator 
$\hat{\cal T}$ gives
\begin{eqnarray}
\hat{\cal T}^{-1} \,\hat{c}^\dagga_{{\bf R},\eta,\uparrow} \,\hat{\cal T} &=& \hat{c}^\dagga_{{\bf R},\eta,\downarrow}, \\
\hat{\cal T}^{-1} \,\hat{c}^\dagga_{{\bf R},\eta,\downarrow} \,\hat{\cal T} &=& - \hat{c}^\dagga_{{\bf R},\eta,\uparrow},
\end{eqnarray}
and includes complex conjugation, i.e., $\hat{\cal T}^{-1} z \hat{\cal T}=z^{*}$ for any complex number $z$. Then, it is possible to define the corresponding 
$2 N_{\rm o} \times 2 N_{\rm o}$ matrices that act on the matrix $h({\bf k})$. The inversion $P=\mathbb{1} \otimes V$ is a symmetry of the model provided 
$P^{-1} \,h({\bf k})\,P = h(-{\bf k})$, while the time-reversal $T=i\,\sigma_y \otimes \mathbb{1}\,K$ (where $K$ takes the complex conjugation) requires 
$T^{-1} h({\bf k}) T = h(-{\bf k})$. The first property implies that it is always possible to choose the eigenstates of $h({\bf k})$ in such a way that 
$u_{-{\bf k},a}=P^\dagger u_{{\bf k},a}$ (provided $-{\bf k} \ne {\bf k} + {\bf G}$), while for $-{\bf k} = {\bf k} + {\bf G}$, the parity operator $P$ 
commutes with $h({\bf k})$ and then it is possible to choose $u_{{\bf k},a}$ as an eigenstate of $P$ with eigenvalue $\xi_a({\bf k})=\pm 1$.  Instead, 
time-reversal symmetry leads to Kramers degeneracy.

\section{The $\mathbb{Z}_2$ many-body marker}\label{sec:marker}

The central object of our analysis is the real space, spin-projected operator~\cite{resta1999,aligia1999}:
\begin{equation}
\hat{Z}_{\sigma}(\delta {\bf k}) = e^{i\delta {\bf k} \cdot \sum_{\bf R} {\bf R} \, \hat{n}^\dagga_{{\bf R},\sigma}},
\end{equation}
where ${\bf R}$ runs over the lattice vectors of the cluster, $\delta {\bf k}$ is a momentum shift, quantized according to the geometry of the cluster (see below),
and $\hat{n}^\dagga_{{\bf R},\sigma}=\sum_{\eta} \hat{c}^\dagger_{{\bf R},\eta,\sigma} \hat{c}^\dagga_{{\bf R},\eta,\sigma}$ is the number operator of electrons on 
site ${\bf R}$ with spin $\sigma$ (summed over the orbital index $\eta$).

Under parity transformation, we have that $\hat{\cal P}^{-1} \,\hat{n}^\dagga_{{\bf R},\sigma} \,\hat{\cal P} = \hat{n}^\dagga_{-{\bf R},\sigma}$ leading to:
\begin{equation}
\hat{\cal P}^{-1} \hat{Z}_{\sigma}(\delta {\bf k}) \hat{\cal P} = e^{-i\delta {\bf k} \cdot \sum_{\bf R} {\bf R} \, \hat{n}^\dagga_{{\bf R},\sigma}} = 
\hat{Z}^\dagger_{\sigma}(\delta {\bf k}).
\end{equation}
Therefore, whenever the parity is not broken, the ground-state average $\langle\Psi|\hat{Z}_{\sigma}(\delta {\bf k})|\Psi\rangle$ is always real, i.e., its phase 
can be either $0$ or $\pi$ (modulo $2\pi$).

In Ref.~\cite{gilardoni2022}, we studied two-dimensional lattice models with $N_{\rm o}=2$, showing that the $\mathbb{Z}_2$ topological character is uniquely 
identified by the (quantized) phase of the marker:
\begin{equation}\label{eq:rho}
\rho_{\sigma} = \frac{\langle\Psi|\hat{Z}_{\sigma}(\delta {\bf k}_1+\delta {\bf k}_2)|\Psi\rangle}
                {\langle\Psi|\hat{Z}_{\sigma}(\delta {\bf k}_1)|\Psi\rangle \,\langle\Psi|\hat{Z}_{\sigma}(\delta {\bf k}_2)|\Psi\rangle},
\end{equation}
where the two elementary momenta are taken as $\delta {\bf k}_n={\bf b}_n/L_n$. For non-interacting models that commute with $S^z$, this marker is simply 
related to the Chern number of the occupied band. Remarkably, even though $\rho_\sigma$ is defined for a given spin projection, its validity is not limited 
to models where the total spin along $z$ is conserved. In fact, we showed that even including a spin-orbit interaction (e.g., the Rashba term in the Kane-Mele 
model) the definition of Eq.~\eqref{eq:rho} allows one to discriminate between trivial and topological phases~\cite{gilardoni2022}. In the following, we show 
that the same marker gives a way to distinguish among trivial, weak-, and strong-topological insulators in three dimensions.

\subsection{The non-interacting limit: models with conserved $S^z$}

Here, we focus on models where the total spin along $z$ commutes with the Hamiltonian~\eqref{eq:hamiltonian}. In this case, the single-particle spectrum 
consists of $N_{\rm o}$ bands per spin projection. In the following, we perform the calculation for the lowest-energy $m$ bands, assuming that a gap separates 
them from the other ones. This circumstance corresponds to an electron density per spin projection $\nu=m/N_{\rm o}$. The many-body ground state $|\Psi\rangle$ 
is the Slater determinant of the single particle orbitals of the occupied bands, which, in this case, is factorized into spin up and down components. The 
$\hat{Z}_\sigma(\delta {\bf k})$ operator acts only on the states of the lowest (occupied) spin-$\sigma$ bands, defining the $m\,N_{\rm c} \times m\,N_{\rm c}$ 
overlap matrix $O$ (see the Supplemental Material of Ref.~\cite{gilardoni2022}):
\begin{equation}\label{eq:overlap}
O_{({\bf q},a),({\bf p},b)} =\langle \psi_{{\bf q},a} |\hat{Z}_\sigma(\delta {\bf k})|\psi_{{\bf p},b}\rangle 
=\delta_{{\bf q},{\bf p}+\delta {\bf k}} \,A_{a,b}^\sigma({\bf q}),
\end{equation}
where the $m \times m$ matrices $A^\sigma({\bf q})$ are defined by:
\begin{equation}
A_{a,b}^\sigma({\bf q})= \sum_{\eta} u^*_{{\bf q},a}(\eta,\sigma) u_{{\bf q}-\delta {\bf k},b}(\eta,\sigma);
\end{equation}
here and in the following, $a$ and $b$ (running from $1$ to $m$) label the occupied spin-$\sigma$ bands. The elementary properties of Slater determinants imply 
that:
\begin{equation}\label{eq:zetadet}
\langle \Psi|\hat{Z}_{\sigma}(\delta {\bf k})|\Psi\rangle = \det \{O \}.
\end{equation}
Given the specific form of the overlap matrix, for an even number of sites in the cluster $N_c$, we have:
\begin{equation}\label{eq:prod}
\langle \Psi|\hat{Z}_\sigma(\delta {\bf k})|\Psi\rangle = \prod_{\bf q} \det \{ A^\sigma(\bf q)\}.
\end{equation}
Inversion symmetry allows us to further simplify this expression. We already pointed out that whenever $-{\bf q} \ne {\bf q} + {\bf G}$ it is always possible 
to choose $u_{-{\bf q},a}=P^\dagger \, u_{{\bf q},a}$, while for $-{\bf q} = {\bf q} + {\bf G}$, $u_{{\bf q},a}$ is an eigenstate of $P$ with eigenvalue 
$\xi_a({\bf q})=\pm 1$. Then, it follows that the overlap matrix $A^\sigma({\bf q})$ satisfies a simple transformation property when both ${\bf q}$ and 
${\bf q}+\delta {\bf k}$ are not invariant under inversion (modulo a reciprocal lattice vector): 
\begin{eqnarray}
A_{b,a}^\sigma(-{\bf q}) &=& \sum_\eta u^*_{-{\bf q},b}(\eta,\sigma) u_{-{\bf q}-\delta {\bf k},a}(\eta,\sigma) \nonumber \\
&=& \sum_\eta \{ [P^\dagger u_{{\bf q},b}](\eta,\sigma) \}^* [P^\dagger u_{{\bf q}+\delta {\bf k},a}](\eta,\sigma) \nonumber \\ 
&=& \sum_\eta u^*_{{\bf q},b}(\eta,\sigma) u_{{\bf q}+\delta {\bf k},a}(\eta,\sigma) \nonumber \\ 
&=& \left[A_{a,b}^\sigma({\bf q}+\delta {\bf k}) \right ]^*.
\label{eq:astar}
\end{eqnarray}
Instead, if ${\bf q}$ remains unchanged under inversion, the transformation property acquires an extra-factor:
\begin{eqnarray}
A_{b,a}^\sigma(-{\bf q}) &=& \sum_\eta u^*_{-{\bf q},b}(\eta,\sigma) u_{-{\bf q}-\delta {\bf k},a}(\eta,\sigma) \nonumber \\
&=& \sum_\eta u^*_{{\bf q},b}(\eta,\sigma) [P^\dagger u_{{\bf q}+\delta {\bf k},a}](\eta,\sigma) \nonumber \\ 
&=& \sum_\eta \{ [P^\dagger u_{{\bf q},b}](\eta,\sigma) \}^* u_{{\bf q}+\delta {\bf k},a}(\eta,\sigma) \nonumber \\ 
&=& \xi_b({\bf q})\,\left [ A_{a,b}^\sigma({\bf q}+\delta {\bf k}) \right ]^*.
\label{eq:astarb}
\end{eqnarray}

Setting $\delta {\bf k}_n={\bf b}_n/L_n$, in a cluster with an even number of sites $L_n$, the factors $\det \{A^\sigma(-{\bf q})\}$ and 
$\det \{ A^\sigma({\bf q}+\delta {\bf k}_n)\}$ always come in pairs. As a consequence, the sign of the product over all wavevectors ${\bf q}$ in 
Eq.~\eqref{eq:prod} is just $\prod_i\delta_i$, with $\delta_i=\prod_a \xi_a({\bf \Gamma}_i)$, where the index $i$ runs over all the TRIM wavevectors 
${\bf \Gamma}_i$ (that are left unchanged under inversion) and the index $a$ runs over all the occupied bands.

In conclusion, when the number of sites of the cluster in each direction $L_j$ is even, the sign of the ground state average of each position operator 
$\hat{Z}_\sigma(\delta {\bf k}_n)$ (with $n=1,2,3$) equals the product of the eight $\delta_i$ and coincides with the $\mathbb{Z}_2$ invariant that 
identifies the strong-topological phase. In addition, the three other (weak) $\mathbb{Z}_2$ topological invariants can be evaluated by taking a cluster 
where one of the $L_j$ (for $j\ne n$) is odd. In this way, four of the special points ${\bf \Gamma}_i$ are no longer present, due to the quantization rules, 
and the sign of the ground-state average of $\hat{Z}_\sigma(\delta {\bf k}_n)$ is just the product of the four remaining $\delta_i$. The trivial phase occurs 
when all topological invariants equal $+1$; the weak-topological phases differ from the trivial insulator because of a negative value of at least one 
(in fact two) of the ground-state averages of $\hat{Z}_\sigma(\delta {\bf k}_n)$ in clusters where one of the $L_j$ (with $j\ne n$) is odd; finally, the 
strong-topological phase appears when the ground state average of $\hat{Z}_\sigma(\delta {\bf k}_n)$ is negative (for any choice of $n$) in clusters with 
even number of sites $L_j$ in all directions.

The ground-state averages of $\hat{Z}_\sigma(\delta {\bf k}_n)$ combine in the definition of the topological marker~\eqref{eq:rho} previously introduced 
in two dimensional models. In the adopted definition, both $\delta {\bf k}_1$ and $\delta {\bf k}_2$ have no components along ${\bf b}_3$, implying that all 
expectation values may be factorized into a product over the third component of ${\bf q}=\sum_{i} q_i {\bf b}_i$ of $q_3$-dependent factors, which immediately 
leads to:
\begin{equation}\label{eq:rhoq3}
\rho_{\sigma} = \prod_{q_3} \, \rho_{\sigma}(q_3);
\end{equation}
here, $\rho_{\sigma}(q_3)$ is the two-dimensional $\mathbb{Z}_2$ marker at fixed $q_3$, i.e., the one evaluated along the $({\bf b}_1,{\bf b}_2)$ planes of the
Brillouin zone at the given $q_3$. In Ref.~\cite{gilardoni2022}, we showed that, in the thermodynamic limit ($N_{\rm c} \to \infty$) and for the case of a single
band, the phase of $\rho_{\sigma}(q_3)$ is just $\pi$ times the spin Chern number of that plane. As previously shown, the sign of this product over all $q_3$'s 
is determined by the special points invariant under inversion symmetry: $q_3=0$ and, when allowed by the quantization condition, $q_3=1/2$. Therefore, only the 
Chern numbers of these two planes contain the relevant information for the identification of a topological insulator. 

In spatially isotropic models, the marker can be equivalently used to identify the three phases: if its sign is positive for both $L_3$ even and odd, the two 
spin Chern numbers at $q_3=0$ and $q_3=1/2$ are even and the insulator is trivial. If $\rho_{\sigma}$ is positive for $L_3$ even while is negative for $L_3$ 
odd, the two spin Chern numbers at $q_3=0$ and $q_3=1/2$ are both odd and the topological insulator is weak. Finally, if $\rho_{\sigma}$ is negative for $L_3$ 
even, the two spin Chern numbers at $q_3=0$ and $q_3=1/2$ have different signs and the topological insulator is strong.

\subsection{The non-interacting limit: The general cases}

In the previous part, we showed that the position operator contains the relevant information required to identify the topological phases in non-interacting 
models where $S^{z}$ commutes with the Hamiltonian. In fact, in these cases, the label $\sigma$ selects one state of the Kramers doublet and allows us to 
evaluate the $\mathbb{Z}_2$ invariant. However, when the total spin along $z$ does not commute with the Hamiltonian, the $\hat{Z}_\sigma(\delta {\bf k})$ 
operator mixes the states of the Kramers doublets, spoiling the mathematical demonstration and making the connection to the eigenvalues of the parity operator 
$\xi_a({\bf \Gamma}_i)$ less transparent.  Still, we consider the spin-projected operator (e.g., with spin up) also in such a case, appealing to adiabatic 
continuity between the model of interest and a model where $S^{z}$ is conserved, provided that the gaps at the TRIM points do not close during the switching 
on of the non-conserving terms. The correctness of this procedure has been already checked in two dimensions~\cite{gilardoni2022} in the simplest case with
$N_{\rm o}=2$, and is now adopted also in three dimensions. The explicit expression for the evaluation of the ground state average of 
$\hat{Z}_\sigma(\delta {\bf k})$ is reported in Appendix~\ref{sec:app}.

\begin{figure}
\includegraphics[width=\columnwidth]{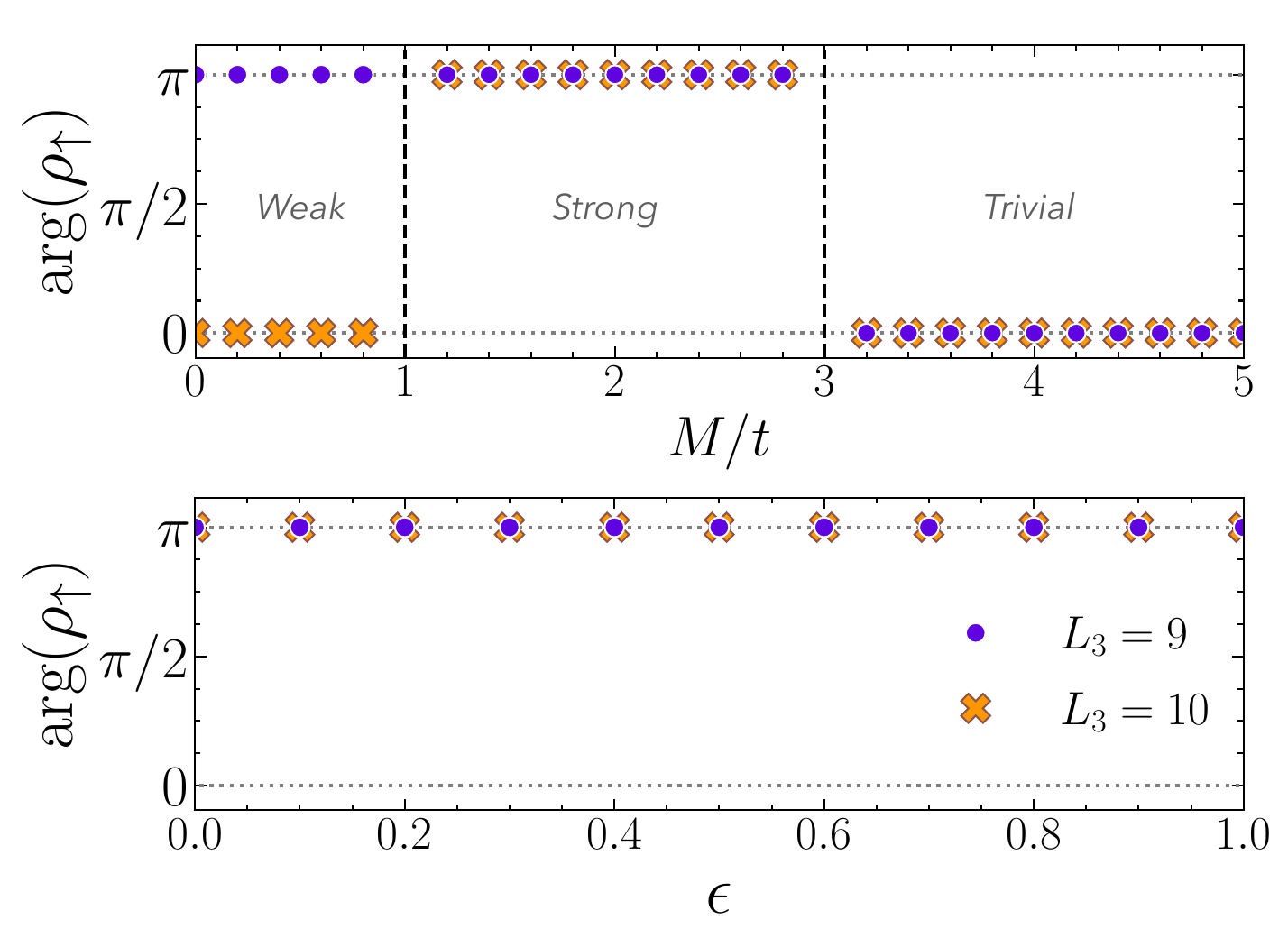}
\caption{\label{fig:rho}
Upper panel: the sign of the many-body marker $\rho_{\uparrow}$ of Eq.~\eqref{eq:rho} for the BHZ (or equivalently the FKM) model, see Eq.~\eqref{eq:bhz_hk} 
for $\lambda/t=0.3$ and $\epsilon=1$ as a function of $M/t$. The results for two clusters with $L_1=L_2=10$ and $L_3=10$ or $9$ are reported. Lower panel: 
the sign of $\rho_{\uparrow}$ for $M/t=2$ (within the strong-topological insulator) as a function of $\epsilon$ (here, $\epsilon=1$ corresponds to the 
original model and $\epsilon=0$ to the model that commutes with the total spin along $z$).}
\end{figure}

\section{Specific models}\label{sec:models}

In the following, we illustrate the results for two specific cases that have been introduced in the past, i.e., the BHZ model on the cubic lattice~\cite{liu2010} and
the Fu-Kane-Mele (FKM) model on the diamond lattice~\cite{fu2007a}. In both cases, we have that $N_{\rm o}=2$, thus simplifing the actual calculations.

\subsection{The BHZ model on the cubic lattice}

The Hamiltonian is translationally invariant but does not conserve the $z$ component of the total spin. The non-interacting case is written in terms of a $4 \times 4$ 
matrix $h({\bf k})$ defined by:
\begin{equation}\label{eq:bhz_hk}
h({\bf k}) =\sum_{\alpha=1}^5 d_{\alpha}({\bf k}) \, \Gamma_{\alpha},
\end{equation}
where $\Gamma_1 = \sigma_z \otimes \tau_x$, $\Gamma_2 = -\mathbb{1} \otimes \tau_y$, $\Gamma_3 = \sigma_x \otimes \tau_x$, $\Gamma_4 = \sigma_y \otimes \tau_x$, and 
$\Gamma_5 = \mathbb{1} \otimes \tau_z$; the coefficients are given by~\cite{amaricci2016}:
\begin{eqnarray}
d_1({\bf k}) &=& \lambda \sin k_x, \nonumber \\
d_2({\bf k}) &=& \lambda \sin k_y, \nonumber \\
d_3({\bf k}) &=& \lambda \sin k_z, \nonumber \\
d_4({\bf k}) &=& 0, \nonumber \\
d_5({\bf k}) &=& M-t(\cos k_x + \cos k_y + \cos k_z). \nonumber
\end{eqnarray}
The model is time-reversal invariant and the inversion symmetry is given by $P=\mathbb{1} \otimes \tau_z$. In addition, the presence of the chiral symmetry 
$C=\sigma_y \otimes \tau_y$, which anticommutes with $h({\bf k})$, gives rise to a particle-hole symmetric spectrum, i.e., two (doubly degenerate) bands with 
$\epsilon({\bf k}) = \pm \sqrt{\sum_{\alpha} d_{\alpha}^2({\bf k})}$. At half filling, a finite gap is present, except for $M/t=\pm 1$ and $M/t=\pm 3$, where Dirac 
points appear at the Fermi level. Taking $M/t \ge 0$, the ground state is known to be a trivial insulator for $M/t>3$, a strong-topological insulator for $1<M/t<3$, 
and a weak-topological insulator for $0 \le M/t<1$.

The $\Gamma_3$ term is responsible for the non-conservation of the $z$ component of the spin. Then, we can generalize the model by introducing an additional parameter 
$\epsilon$ in front of this term, i.e., $d_3({\bf k})$: $d_3({\bf k}) \to \epsilon \, d_3({\bf k})$, and study how the marker evolves from the original model with 
$\epsilon=1$ to the model that commutes with the total spin along $z$ (with $\epsilon=0$). Notice that for $\epsilon=0$, the spectrum is gapless for $0 \le M/t \le 3$, 
but the Dirac points are shifted to incommensurate values fo $k_z$. Still, the calculation of the sign of $\rho_{\sigma}$ is not affected by this fact, since it depends 
only on specific planes in the Brilloiun zone. We evaluate the marker $\rho_{\uparrow}$ for $L_1=L_2=10$ and either $L_3=10$ or $L_3=9$ (its sign depending only on
the parity of $L_i$, in agreement with the previous analysis, with no size effects). The results are shown in Fig.~\ref{fig:rho}, for $\epsilon=1$ varying $M/t$ and 
for $M/t=2$ varying $\epsilon$. In the former case, for $M/t>3$, the the marker is {\it positive} for both values of $L_3$ (confirming the fact that the ground state 
is a trivial insulator); conversely, for $1<M/t<3$, the marker is {\it negative} for both values of $L_3$ (corresponding to a strong-topological insulator); finally, 
for $0 \le M/t<1$, $\rho_{\uparrow}$ is {\it positive} for $L_3=10$ and {\it negative} for $L_3=9$ (indicating a weak-topological insulator).

\subsection{The FKM model on the diamond lattice}

The FKM model is defined on the diamond lattice (a face-centered cubic lattice with a two-site basis) and is given by the same expression of Eq.~\eqref{eq:bhz_hk} 
with $\Gamma_1 = \mathbb{1}\otimes\tau_x$, $\Gamma_2 = \mathbb{1}\otimes\tau_y$, $\Gamma_3 = \sigma_x\otimes\tau_z$, $\Gamma_4 = \sigma_y\otimes\tau_z$, and
$\Gamma_5 = \sigma_z\otimes\tau_z$. The coefficients are given by~\cite{fu2007a}
\begin{widetext}
\begin{eqnarray}
d_1({\bf k}) &=& M + t(\cos x_1 + \cos x_2 + \cos x_3), \nonumber \\
d_2({\bf k}) &=& t(\sin x_1 + \sin x_2 + \sin x_3), \nonumber \\
d_3({\bf k}) &=& \lambda \left [ \sin x_2 - \sin x_3 - \sin (x_2-x_1) + \sin (x_3-x_1) \right ], \nonumber \\
d_4({\bf k}) &=& \lambda \left [ \sin x_3 - \sin x_1 - \sin (x_3-x_2) + \sin (x_1-x_2) \right ], \nonumber \\
d_5({\bf k}) &=& \lambda \left [ \sin x_1 - \sin x_2 - \sin (x_1-x_3) + \sin (x_2-x_3) \right ], \nonumber
\end{eqnarray}
\end{widetext}
where $M=1+\delta t$ (as defined in the original work~\cite{fu2007a}) and $x_i = {\bf k} \cdot {\bf a}_i$. As in the BHZ model, the Hamiltonian is time-reversal 
invariant. In addition, the inversion symmetry is given by $P=\mathbb{1} \otimes \tau_x$. Also in this case there are two (doubly degenerate) bands with 
$\epsilon({\bf k}) = \pm \sqrt{\sum_{\alpha} d_{\alpha}^2({\bf k})}$. Within the present choice of the ``mass'' term $M$, the FKM and BHZ models share the 
same ground-state phase diagram. In this case, the limit in which the Hamiltonian commutes with the total spin along $z$ is obtained by introducing the parameter 
$\epsilon$ in front of both $d_3({\bf k})$ and $d_4({\bf k})$ terms. The results for the sign of the marker coincide with the ones reported before in the BHZ
model, see Fig.~\ref{fig:rho}.

\section{Conclusions}\label{sec:concl}

In summary, we have shown that the local many-body marker of Eq.~\eqref{eq:rho}, which we introduced to distinguish $\mathbb{Z}_2$ topological insulators in two 
spatial dimensions~\cite{gilardoni2022}, can be also used to discriminate among trivial, weak-, and strong-topological insulators in three dimensions, in presence 
of inversion symmetry. Explicit calculations in non-interacting limits have been reported. In particular, the presence of the inversion symmetry forces the marker 
to be real and the nature of the insulator may be extracted from its sign. Within non-interacting models in which $S^{z}$ is conserved, the sign is determined by 
the parity eigenvalues of the occupied orbitals at time-reversal invariant momenta in the Brillouin zone, thus reproducing the results obtained in Ref.~\cite{fu2007b}. 
For general models, with non conserved $S^{z}$, this correspondence is no longer transparent, but it is expected to remain whenever it is possible to adiabatically 
switch-off the non-conserving terms in the Hamiltonian without closing the gap at the TRIM points.

Most importantly, since our marker is not based upon the single-particle picture, it can be employed also to assess the properties of electronic models in the presence 
of electron-electron interactions, by computing ground-state expectation values, which are accessible within quantum Monte Carlo approaches, including the variational
one (as recently done for the BHZ model with Hubbard and nearest-neighbor density-density interactions~\cite{favata2025}). Still, these kind of calculations in three
dimensions require intensive numerical efforts and, therefore, are left for future studies.

\section*{Acknowledgements}
We thank A. Marrazzo, L. Balents, and G. Sangiovanni for useful discussions on topological markers. We also are indebted to R. Favata for discussions and for preparing
the figure. 

\appendix
\section{Details of the general case}\label{sec:app}

Here, we report the derivation of the explicit expression of the ground-state average of $\hat Z_\uparrow(\delta{\bf k})$ in models where $S^{z}$ is non conserved. 
The single-particle orbitals are defined in Eq.~\eqref{eq:bloch}. As before, we consider a model with $N_o$ orbitals per Bravais lattice site, implying $2N_o$ 
bands, $2m$ of which are occupied. Now, the overlap matrix introduced in Eq.~\eqref{eq:overlap} has dimension $2mN_c\times 2mN_c$ and is given by:
\begin{equation}
O_{({\bf q},a),({\bf p},b)} = \delta_{{\bf q},{\bf p}+\delta{\bf k}} \, A_{a,b}^\uparrow({\bf q}) -\delta_{{\bf q},{\bf p}} \, B_{a,b}^\downarrow({\bf q}),
\end{equation}
where the two $2m\times 2m$ matrices $A^\uparrow({\bf q})$ and $B^\downarrow({\bf q})$ are defined by
\begin{eqnarray}
A_{a,b}^\uparrow({\bf q}) &=&  \sum_\eta u^*_{{\bf q},a}(\eta,\uparrow)\,u_{{\bf q}-\delta {\bf k},b}(\eta,\uparrow), \\
B_{a,b}^\downarrow({\bf q}) &=&  -\sum_\eta u^*_{{\bf q},a}(\eta,\downarrow)\,u_{{\bf q},b}(\eta,\downarrow).
\end{eqnarray}

In order to evaluate the determinant of the overlap matrix, it is convenient to investigate the corresponding eigenvalue problem:
\begin{equation}\label{eq:eigenxq}
A^\uparrow({\bf q}) \, x({\bf q}-\delta {\bf k}) - B^\downarrow({\bf q}) \, x({\bf q}) = \lambda \, x({\bf q}),
\end{equation}
which defines the recurrence relation:
\begin{equation}
x({\bf q}-\delta{\bf k}) = \left \{ \lambda [A^\uparrow({\bf q})]^{-1}+ [A^\uparrow({\bf q})]^{-1} \, B^\downarrow({\bf q}) \right \} \, x({\bf q}).
\end{equation}
Let us now take $\delta {\bf k} = \delta {\bf k}_n={\bf b}_n/L_n$. Then, the eigenvalue problem reduces to a set of $2mL_n\times 2mL_n$ blocks, 
since every value of ${\bf q}$ only couples to ${\bf q}-j \delta {\bf k}_n$, with $j=0,\dots,L_n-1$, and $L_n \delta{ \bf k}_n$ equals ${\bf b}_n$, 
a vector of the reciprocal lattice. Therefore, starting from a vector ${\bf q}_{\perp}$, which has no component along ${\bf b}_n$ (i.e., such that 
${\bf q} \cdot {\bf a}_n =0$) and iterating $L_n-1$ times, we get:
\begin{widetext}
\begin{equation}\label{eq:eig}
x({\bf q}_{\perp}-L_n \delta {\bf k}_n) = \left \{ \prod_{j=0}^{L_n-1} \left \{ \lambda [A^\uparrow({\bf q}_{\perp}-j \delta {\bf k}_n)]^{-1} + 
[A^\uparrow({\bf q}_{\perp}-j \delta {\bf k}_n)]^{-1}\, B^\downarrow({\bf q}_{\perp}-j \delta {\bf k}_n) \right \} \right \} \, x({\bf q}_{\perp}),
\end{equation}
where the product is ordered in $j$. Since $x({\bf q}-L_n \delta {\bf k}_n) \equiv x({\bf q})$, this relation implies that the $2m\times 2m$ matrix
\begin{equation}
M({\bf q}_{\perp}) = \prod_{j=0}^{L_n-1} \left \{ \lambda [A^\uparrow({\bf q}_{\perp}-j \delta {\bf k}_n)]^{-1}+ 
[A^\uparrow({\bf q}_{\perp}-j \delta {\bf k}_n)]^{-1} \, B^\downarrow({\bf q}_{\perp}-j \delta {\bf k}_n) \right \}
\end{equation}
\end{widetext}
has an eigenvalue equal to one. Thus, the corresponding secular equation is $\det\{ M({\bf q}_{\perp}) - {\mathbb{1}}\}= 0$ for each choice of 
${\bf q}_{\perp}$. As a result, the overlap matrix is a block matrix (each block being a $2mL_n\times 2mL_n$ matrix), and its determinant is the 
product over ${\bf q}_{\perp}$ of the determinants of each block. The determinant of every block can be easily computed: The secular equation is 
a polynomial of degree $2mL_n$ in $\lambda$, e.g., $C\,(\lambda-\lambda_1) \dots (\lambda-\lambda_{2mL_n})$, where $C$ is a constant and $\lambda_i$ 
are the $2mL_n$ eigenvalues.  From the factorization property, it follows that the determinant of each block matrix equals the ratio between the 
term of zero degree in $\lambda$ and the term of degree $2mL_n$ in the secular equation: 
\begin{widetext}
\begin{equation}\label{eq:Rfactor}
R({\bf q}_{\perp}) = \frac{\det\{\prod_{j=0}^{L_n-1} [A^\uparrow({\bf q}_{\perp}-j \delta {\bf k}_n)]^{-1} \, 
B^\downarrow({\bf q}_{\perp}-j \delta {\bf k}_n) -{\mathbb {1}}\}} {\det\{\prod_{j=0}^{L_n-1}[A^\uparrow({\bf q}_{\perp}-j \delta {\bf k}_n)]^{-1}\}}.
\end{equation}
\end{widetext}
Finally, the determinant of the full overlap matrix is the product over ${\bf q}_{\perp}$ of these factors:
\begin{equation}\label{eq:z1}
\langle \Psi| \hat Z_\uparrow(\delta {\bf k}_n) |\Psi \rangle = \prod_{{\bf q}_{\perp}} \, R({\bf q}_{\perp}).
\end{equation}

Inversion symmetry allows us to further simplify this expression. Let us consider the pair of terms corresponding to ${\bf q}_{\perp}$ and 
$-{\bf q}_{\perp}$ (modulo a reciprocal lattice vector) in the product of Eq.~\eqref{eq:z1}, assuming that they differ from the TRIM (i.e., that 
${\bf q}_{\perp} \ne {\bf \Gamma}_i$). From Eq.~\eqref{eq:astar}, which remains valid also in non spin conserving models, it follows that: 
\begin{eqnarray}
&& \det A^\uparrow(-{\bf q}_{\perp}-j \delta {\bf k}_n) = \det \{ [A^\uparrow({\bf q}_{\perp}+(j+1) \delta {\bf k}_n)]^\dagger \} \nonumber \\
&&= [\det A^\uparrow({\bf q}_{\perp}-(L_n-1-j) \delta {\bf k}_n)]^*.
\end{eqnarray}
Therefore, the product of the terms ${\bf q}_{\perp}$ and $-{\bf q}_{\perp}$ in the denominator of Eq.~\eqref{eq:Rfactor} always gives a positive number. 

A similar argument holds also for the expression in the numerator, although the derivation is less straightforward: 
\begin{widetext}
\begin{equation}
\det \left \{ \prod_{j=0}^{L_n-1} [A^\uparrow(-{\bf q}_{\perp}-j \delta {\bf k}_n)]^{-1} \, 
B^\downarrow(-{\bf q}_{\perp}-j \delta {\bf k}_n) -\mathbb{1}\right \}
=
\left [ \det \left \{\prod_{j=0}^{L_n-1} [A^\uparrow({\bf q}_{\perp}-j \delta {\bf k}_n)]^{-1} \, 
B^\downarrow({\bf q}_{\perp}-j \delta {\bf k}_n) -\mathbb{1}\right \}\right ]^*.
\end{equation}
\end{widetext}
Therefore, the sign of $\langle \Psi| \hat Z_\uparrow(\delta {\bf k}_n)|\Psi \rangle$ in Eq.~\eqref{eq:z1} is determined by the factors corresponding to momenta 
${\bf q}_{\perp}$ equal to the TRIM ${\bf \Gamma}_i$ that are orthogonal to ${\bf a}_n$. 

In the simplest case of two-band models, it is easy to show that the sign of $\langle \Psi| \hat Z_\uparrow(\delta {\bf k}_n)|\Psi \rangle$ is again given by the 
product of the parity eigenvalues at the TRIMs, as expected. This fact requires that the gap at the TRIMs does not close during the adiabatic switching-on of the 
terms that do not conserve $S^z$. 

\bibliography{biblio}

\end{document}